\documentclass[12pt,a4paper]{article}

\usepackage{ifthen} 
\newboolean{pdflatex}
\setboolean{pdflatex}{true} 

\newboolean{articletitles}
\setboolean{articletitles}{true} 

\newboolean{uprightparticles}
\setboolean{uprightparticles}{false} 

\def\paperauthors{LHCb collaboration} 
\def\paperasciititle{Stringent limits on CPT- and Lorentz-invariance violation from B_s^0 meson decays} 
\def\papertitle{Stringent limits on $CPT$- and Lorentz-invariance violation from $B_s^0$ meson decays} 
\def\paperkeywords{{High Energy Physics}, {LHCb}} 
\def\papercopyright{\the\year\ CERN for the benefit of the LHCb collaboration} 
\def\paperlicence{CC BY 4.0 licence}
\def\paperlicenceurl{https://creativecommons.org/licenses/by/4.0/}

\newif\ifEnableSectionTOCLinks
\EnableSectionTOCLinksfalse 

\usepackage[top=1in, bottom=1.25in, left=1in, right=1in]{geometry}

\usepackage{microtype}
\usepackage{lineno}  
\usepackage{xspace} 
\usepackage{caption} 

\usepackage{graphicx}  
\usepackage{color}
\usepackage{colortbl}
\graphicspath{{./figs/}} 

\usepackage{amsmath} 
\usepackage{amssymb}
\usepackage{amsfonts}
\usepackage{upgreek} 

\usepackage[normalem]{ulem} 

\newcommand*\patchAmsMathEnvironmentForLineno[1]{%
\expandafter\let\csname old#1\expandafter\endcsname\csname #1\endcsname
\expandafter\let\csname oldend#1\expandafter\endcsname\csname
end#1\endcsname
 \renewenvironment{#1}%
   {\linenomath\csname old#1\endcsname}%
   {\csname oldend#1\endcsname\endlinenomath}%
}
\newcommand*\patchBothAmsMathEnvironmentsForLineno[1]{%
  \patchAmsMathEnvironmentForLineno{#1}%
  \patchAmsMathEnvironmentForLineno{#1*}%
}
\AtBeginDocument{%
\patchBothAmsMathEnvironmentsForLineno{equation}%
\patchBothAmsMathEnvironmentsForLineno{align}%
\patchBothAmsMathEnvironmentsForLineno{flalign}%
\patchBothAmsMathEnvironmentsForLineno{alignat}%
\patchBothAmsMathEnvironmentsForLineno{gather}%
\patchBothAmsMathEnvironmentsForLineno{multline}%
\patchBothAmsMathEnvironmentsForLineno{eqnarray}%
}

\usepackage[pdftex,
            pdfauthor={\paperauthors},
            pdftitle={\paperasciititle},
            pdfkeywords={\paperkeywords}]{hyperref}
\usepackage{hyperxmp}
\hypersetup{
    pdfcopyright={Copyright (C) \papercopyright},
    pdflicenseurl={\paperlicenceurl}
}

\usepackage[colorinlistoftodos,textsize=scriptsize]{todonotes}

\usepackage[bottom,flushmargin,hang,multiple]{footmisc}

\usepackage[all]{hypcap} 

\usepackage{xspace} 
\usepackage{upgreek}

\def\lhcb   {\mbox{LHCb}\xspace}

\def\dzero  {\mbox{D0}\xspace}

\def\lhc    {\mbox{LHC}\xspace}

\def\rich   {RICH\xspace}

\def\MagUp {\mbox{\em Mag\kern -0.05em Up}\xspace}

\ifdefined\ifuprightparticles
\else
\newboolean{uprightparticles}
\setboolean{uprightparticles}{false} 
\fi

\ifthenelse{\boolean{uprightparticles}}%
{

 \def\Peta        {\ensuremath{\upeta}\xspace}

 \def\Ppi         {\ensuremath{\uppi}\xspace}

 \def\Ppsi        {\ensuremath{\uppsi}\xspace}                 
 \def\Pomega      {\ensuremath{\upomega}\xspace}                 

 \def\PDelta      {\ensuremath{\Delta}\xspace}                 
 \def\PXi         {\ensuremath{\Xi}\xspace}                 
 \def\PLambda     {\ensuremath{\Lambda}\xspace}                 
 \def\PSigma      {\ensuremath{\Sigma}\xspace}                 
 \def\POmega      {\ensuremath{\Omega}\xspace}                 
 \def\PUpsilon    {\ensuremath{\Upsilon}\xspace}
 \let\oldPi\Pi
 \def\PPi         {\ensuremath{\oldPi}\xspace}

 \def\PB      {\ensuremath{\mathrm{B}}\xspace}                 
 \def\PD      {\ensuremath{\mathrm{D}}\xspace}                 
 \def\PJ      {\ensuremath{\mathrm{J}}\xspace}                 
 \def\PK      {\ensuremath{\mathrm{K}}\xspace}                 
 \def\Pb      {\ensuremath{\mathrm{b}}\xspace}                 
 \def\Pc      {\ensuremath{\mathrm{c}}\xspace}                 
 \def\Pd      {\ensuremath{\mathrm{d}}\xspace}                 

 \def\Pp      {\ensuremath{\mathrm{p}}\xspace}                 

 \def\Ps      {\ensuremath{\mathrm{s}}\xspace}

 \def\thebaroffset{0.0em}
}
{

 \def\Peta        {\ensuremath{\eta}\xspace}

 \def\Ppi         {\ensuremath{\pi}\xspace}

 \def\Ppsi        {\ensuremath{\psi}\xspace}                 
 \def\Pomega      {\ensuremath{\omega}\xspace}                 
 \mathchardef\PDelta="7101
 \mathchardef\PXi="7104
 \mathchardef\PLambda="7103
 \mathchardef\PSigma="7106
 \mathchardef\POmega="710A
 \mathchardef\PUpsilon="7107
 \mathchardef\PPi="7105
 \def\PB      {\ensuremath{B}\xspace}                 
 \def\PD      {\ensuremath{D}\xspace}                 
 \def\PJ      {\ensuremath{J}\xspace}                 
 \def\PK      {\ensuremath{K}\xspace}                 
 \def\Pb      {\ensuremath{b}\xspace}                 
 \def\Pc      {\ensuremath{c}\xspace}                 
 \def\Pd      {\ensuremath{d}\xspace}                 

 \def\Pp      {\ensuremath{p}\xspace}                 

 \def\Ps      {\ensuremath{s}\xspace}

 \def\thebaroffset{0.18em}
}
\newcommand{\offsetoverline}[2][\thebaroffset]{\kern #1\overline{\kern -#1 #2}}%

\makeatletter
\ifcase \@ptsize \relax
  \newcommand{\miniscule}{\@setfontsize\miniscule{4}{5}}
\or
  \newcommand{\miniscule}{\@setfontsize\miniscule{5}{6}}
\or
  \newcommand{\miniscule}{\@setfontsize\miniscule{5}{6}}
\fi
\makeatother

\DeclareRobustCommand{\optbar}[1]{\shortstack{{\miniscule (\rule[.5ex]{1.25em}{.18mm})}
  \\ [-.7ex] $#1$}}

\def\dquark    {{\ensuremath{\Pd}}\xspace}

\def\squark    {{\ensuremath{\Ps}}\xspace}
\def\squarkbar {{\ensuremath{\overline \squark}}\xspace}

\def\cquark    {{\ensuremath{\Pc}}\xspace}

\def\bquark    {{\ensuremath{\Pb}}\xspace}
\def\bquarkbar {{\ensuremath{\overline \bquark}}\xspace}

\def\pion   {{\ensuremath{\Ppi}}\xspace}

\def\pip    {{\ensuremath{\pion^+}}\xspace}
\def\pim    {{\ensuremath{\pion^-}}\xspace}
\def\pipm   {{\ensuremath{\pion^\pm}}\xspace}

\def\kaon    {{\ensuremath{\PK}}\xspace}

\def\KorKbar {\kern \thebaroffset\optbar{\kern -\thebaroffset \PK}{}\xspace}

\def\Kp      {{\ensuremath{\kaon^+}}\xspace}
\def\Km      {{\ensuremath{\kaon^-}}\xspace}
\def\Kpm     {{\ensuremath{\kaon^\pm}}\xspace}

\def\D       {{\ensuremath{\PD}}\xspace}

\def\DorDbar {\kern \thebaroffset\optbar{\kern -\thebaroffset \PD}\xspace}

\def\Dp      {{\ensuremath{\D^+}}\xspace}
\def\Dm      {{\ensuremath{\D^-}}\xspace}

\def\Dmp     {{\ensuremath{\D^\mp}}\xspace}
\def\DpDm    {\ensuremath{\Dp {\kern -0.16em \Dm}}\xspace}

\def\Dsp     {{\ensuremath{\D^+_\squark}}\xspace}
\def\Dsm     {{\ensuremath{\D^-_\squark}}\xspace}

\def\Dsmp    {{\ensuremath{\D^{\mp}_\squark}}\xspace}

\def\Dssm    {{\ensuremath{\D^{*-}_\squark}}\xspace}

\def\B       {{\ensuremath{\PB}}\xspace}
\def\Bbar    {{\ensuremath{\offsetoverline{\PB}}}\xspace}

\def\BorBbar {\kern \thebaroffset\optbar{\kern -\thebaroffset \PB}\xspace}

\def\Bd      {{\ensuremath{\B^0}}\xspace}

\def\BdorBdbar {\kern \thebaroffset\optbar{\kern -\thebaroffset \Bd}\xspace}
\def\Bu      {{\ensuremath{\B^+}}\xspace}

\def\Bs      {{\ensuremath{\B^0_\squark}}\xspace}
\def\Bsb     {{\ensuremath{\Bbar{}^0_\squark}}\xspace}
\def\BsorBsbar {\kern \thebaroffset\optbar{\kern -\thebaroffset \Bs}\xspace}

\def\jpsi     {{\ensuremath{{\PJ\mskip -3mu/\mskip -2mu\Ppsi}}}\xspace}

\def\Y#1S{\ensuremath{\PUpsilon{(#1S)}}\xspace}

\def\proton      {{\ensuremath{\Pp}}\xspace}

\def\Lbar        {{\ensuremath{\offsetoverline{\PLambda}}}\xspace}
\def\LorLbar     {\kern \thebaroffset\optbar{\kern -\thebaroffset \PLambda}\xspace}

\def\Lcbar       {{\ensuremath{\Lbar{}^-_\cquark}}\xspace}

\def\Lbbar        {{\ensuremath{\Lbar{}^0_\bquark}}\xspace}

\newcommand{\decay}[2]{\ensuremath{\mathinner{#1\!\to #2}}\xspace}

\def\to                 {\ensuremath{\rightarrow}\xspace}

\def\CP                {{\ensuremath{C\!P}}\xspace}
\def\CPT               {{\ensuremath{C\!PT}}\xspace}

\newcommand{\dms}{{\ensuremath{\Delta m_{\squark}}}\xspace}
\newcommand{\dmd}{{\ensuremath{\Delta m_{\dquark}}}\xspace}

\newcommand{\DGs}{{\ensuremath{\Delta\Gamma_{\squark}}}\xspace}
\newcommand{\DGd}{{\ensuremath{\Delta\Gamma_{\dquark}}}\xspace}
\newcommand{\Gs}{{\ensuremath{\Gamma_{\squark}}}\xspace}
\newcommand{\Gd}{{\ensuremath{\Gamma_{\dquark}}}\xspace}

\def\BsToJPsiPhi  {\decay{\Bs}{\jpsi\phi}}

\def\AT#1     {\ensuremath{A_{\mathrm{T}}^{#1}}\xspace}           

\def\C#1      {\ensuremath{\mathcal{C}_{#1}}\xspace}                       
\def\Cp#1     {\ensuremath{\mathcal{C}_{#1}^{'}}\xspace}                    
\def\Ceff#1   {\ensuremath{\mathcal{C}_{#1}^{\mathrm{(eff)}}}\xspace}        
\def\Cpeff#1  {\ensuremath{\mathcal{C}_{#1}^{'\mathrm{(eff)}}}\xspace}       
\def\Ope#1    {\ensuremath{\mathcal{O}_{#1}}\xspace}                       
\def\Opep#1   {\ensuremath{\mathcal{O}_{#1}^{'}}\xspace}                    

\newcommand{\aunit}[1]{\ensuremath{\text{\,#1}}}       

\newcommand{\tev}{\aunit{Te\kern -0.1em V}\xspace}
\newcommand{\gev}{\aunit{Ge\kern -0.1em V}\xspace}
\newcommand{\mev}{\aunit{Me\kern -0.1em V}\xspace}
\newcommand{\kev}{\aunit{ke\kern -0.1em V}\xspace}
\newcommand{\ev}{\aunit{e\kern -0.1em V}\xspace}
 
\newcommand{\mevc}{\ensuremath{\aunit{Me\kern -0.1em V\!/}c}\xspace}
\newcommand{\gevc}{\ensuremath{\aunit{Ge\kern -0.1em V\!/}c}\xspace}
\newcommand{\mevcc}{\ensuremath{\aunit{Me\kern -0.1em V\!/}c^2}\xspace}
\newcommand{\gevcc}{\ensuremath{\aunit{Ge\kern -0.1em V\!/}c^2}\xspace}

\def\cm   {\aunit{cm}\xspace}

\def\fb   {\ensuremath{\aunit{fb}}\xspace}
\def\invfb   {\ensuremath{\fb^{-1}}\xspace}

\def\ps   {\ensuremath{\aunit{ps}}\xspace}

\def\gsim{{~\raise.15em\hbox{$>$}\kern-.85em
          \lower.35em\hbox{$\sim$}~}\xspace}
\def\lsim{{~\raise.15em\hbox{$<$}\kern-.85em
          \lower.35em\hbox{$\sim$}~}\xspace}

\newcommand{\Real}{\ensuremath{\mathcal{R}e}\xspace}
\newcommand{\Imag}{\ensuremath{\mathcal{I}m}\xspace}

\def\sPlot{\mbox{\em sPlot}\xspace}

\def\pt         {\ensuremath{p_{\mathrm{T}}}\xspace}

\def\evtgen     {\mbox{\textsc{EvtGen}}\xspace}

\def\geant      {\mbox{\textsc{Geant4}}\xspace}

\def\photos     {\mbox{\textsc{Photos}}\xspace}

\def\pythia     {\mbox{\textsc{Pythia}}\xspace}

\def\tell1  {TELL1\xspace}
\def\ukl1   {UKL1\xspace}

\newcommand{\lhcborcid}[1]{\href{https://orcid.org/#1}{\hspace*{0.1em}\raisebox{-0.45ex}{\includegraphics[width=1em]{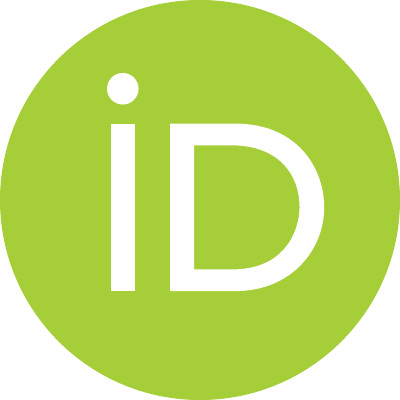}}}}

\hypersetup{
  colorlinks   = true, 
  urlcolor     = blue, 
  linkcolor    = blue, 
  citecolor    = red   
}

\ifEnableSectionTOCLinks
    \usepackage[explicit]{titlesec} 
    
    \let\oldcontentsline\contentsline
    \renewcommand

    \titleformat{\section}{\normalfont\Large\bf}{\hyperlink{tocsection.\thesection}{{\thesection} \parbox[t]{\dimexpr\textwidth-1pc}{#1}}}{1pc}{}

    \titleformat{\subsection}{\normalfont\bf}{\hyperlink{tocsubsection.\thesubsection}{{\thesubsection} \parbox[t]{\dimexpr\textwidth-1pc}{#1}}}{1pc}{}

    \titleformat{name=\section,numberless}[display]{}{}{0pt}{\normalfont\Huge\bfseries #1}
\fi

\usepackage{cite} 
\usepackage{LHCb/mciteplus}
\makeatletter
\g@addto@macro\bfseries{\boldmath}
\makeatother

\def\schr    {\mbox{Schr\"{o}dinger}\xspace}
\def\Dsmpip      {{\ensuremath{\Dsm\pip}}\xspace}
\def\BsToDsPip  {\decay{\Bs}{\Dsmpip}}

\def\z                {{\ensuremath{z}}\xspace}
\def\Rez           {{\ensuremath{\Real(\z)}}\xspace}
\def\Imz           {{\ensuremath{\Imag(\z)}}\xspace}
\def\Damu        {{\ensuremath{\Delta a_{\mu}}\xspace}}
\def\BsL      {{\ensuremath{\B^0_{\squark\mathrm{L}}}}\xspace}
\def\BsH      {{\ensuremath{\B^0_{\squark\mathrm{H}}}}\xspace}
\def\mHL      {{\ensuremath{m_{\mathrm{H,L}}}}\xspace}
\def\msH      {{\ensuremath{m_{\squark\mathrm{H}}}}\xspace}
\def\msL      {{\ensuremath{m_{\squark\mathrm{L}}}}\xspace}
\def\GHL      {{\ensuremath{\Gamma_{\mathrm{H,L}}}}\xspace}
\def\GsH      {{\ensuremath{\Gamma_{\squark\mathrm{H}}}}\xspace}
\def\GsL      {{\ensuremath{\Gamma_{\squark\mathrm{L}}}}\xspace}
\newcommand{\delatms}{{\ensuremath{\delta m_{\squark}}}\xspace}
\newcommand{\dGs}{{\ensuremath{\delta \Gamma_{\squark}}}\xspace}
\def\Dsppim      {{\ensuremath{\Dsp\pim}}\xspace}
\def\BsToDsPim  {\decay{\Bs}{\Dsppim}}

\def\KmKpPim      {{\ensuremath{\Km\Kp\pim}}\xspace}
\def\PimPipPim      {{\ensuremath{\pim\pip\pim}}\xspace}
\def\Damu   {{\ensuremath{\Delta a_{\mu}}}\xspace}
\def\Daze        {{\ensuremath{\Delta a_0}}\xspace}
\def\Dax        {{\ensuremath{\Delta a_X}}\xspace}
\def\Day        {{\ensuremath{\Delta a_Y}}\xspace}
\def\Daz        {{\ensuremath{\Delta a_Z}}\xspace}
\def\Dat        {{\ensuremath{\Daze - 0.38 \Daz}}\xspace}
\def\Das        {{\ensuremath{0.38 \Daze + \Daz}}\xspace}
\def\Dap        {{\ensuremath{\Delta a_{\parallel}}}\xspace}
\def\Dav        {{\ensuremath{\Delta a_{\perp}}}\xspace}

\def\DatResult           {(+0.62 \pm 0.71 \pm 0.32) \times 10^{-14} \gev}
\def\DasResult           {(+0.16 \pm 0.22 \pm 0.07) \times 10^{-12} \gev}
\def\DaxResult           {(-0.11 \pm 1.21 \pm 0.53) \times 10^{-14} \gev}
\def\DayResult           {(+0.78 \pm 1.16 \pm 0.51) \times 10^{-14} \gev}
\def\RezResult           {0.018 \pm 0.016 \pm 0.003}
\def\ImzResult           {0.003 \pm 0.004 \pm 0.002}
\def\qf           {{\ensuremath{\zeta}}\xspace}

\def\Dsmpip      {{\ensuremath{\Dsm\pip}}\xspace}
\def\Dsppim      {{\ensuremath{\Dsp\pim}}\xspace}

\def\BdToDPi  {\decay{\Bd}{\Dm\pip}}
\def\BdToDmpPipm  {\decay{\Bd}{\Dmp\pipm}}

\def\LbToLcPib  {\decay{\Lbbar}{\Lcbar\pip}}

\def\BsToDsstPi  {\decay{\Bs}{\Dssm\pip}}
\def\DssmToDsgamma  {\decay{\Dssm}{\Dsm\gamma}}
\def\BdToDsPi  {\decay{\Bd}{\Dsm\pip}}
\def\BsToDsK  {\decay{\Bs}{\Dsmp\Kpm}}
\def\BuToJpisKp  {\decay{\Bu}{\jpsi\Kp}}
\usepackage{longtable} 

\begin{document}
\renewcommand{\thefootnote}{\fnsymbol{footnote}}
\setcounter{footnote}{1}


\begin{titlepage}
\pagenumbering{roman}

\vspace*{-1.5cm}
\centerline{\large EUROPEAN ORGANIZATION FOR NUCLEAR RESEARCH (CERN)}
\vspace*{1.5cm}
\noindent
\begin{tabular*}{\linewidth}{lc@{\extracolsep{\fill}}r@{\extracolsep{0pt}}}
\ifthenelse{\boolean{pdflatex}}
{\vspace*{-1.5cm}\mbox{\!\!\!\includegraphics[width=.14\textwidth]{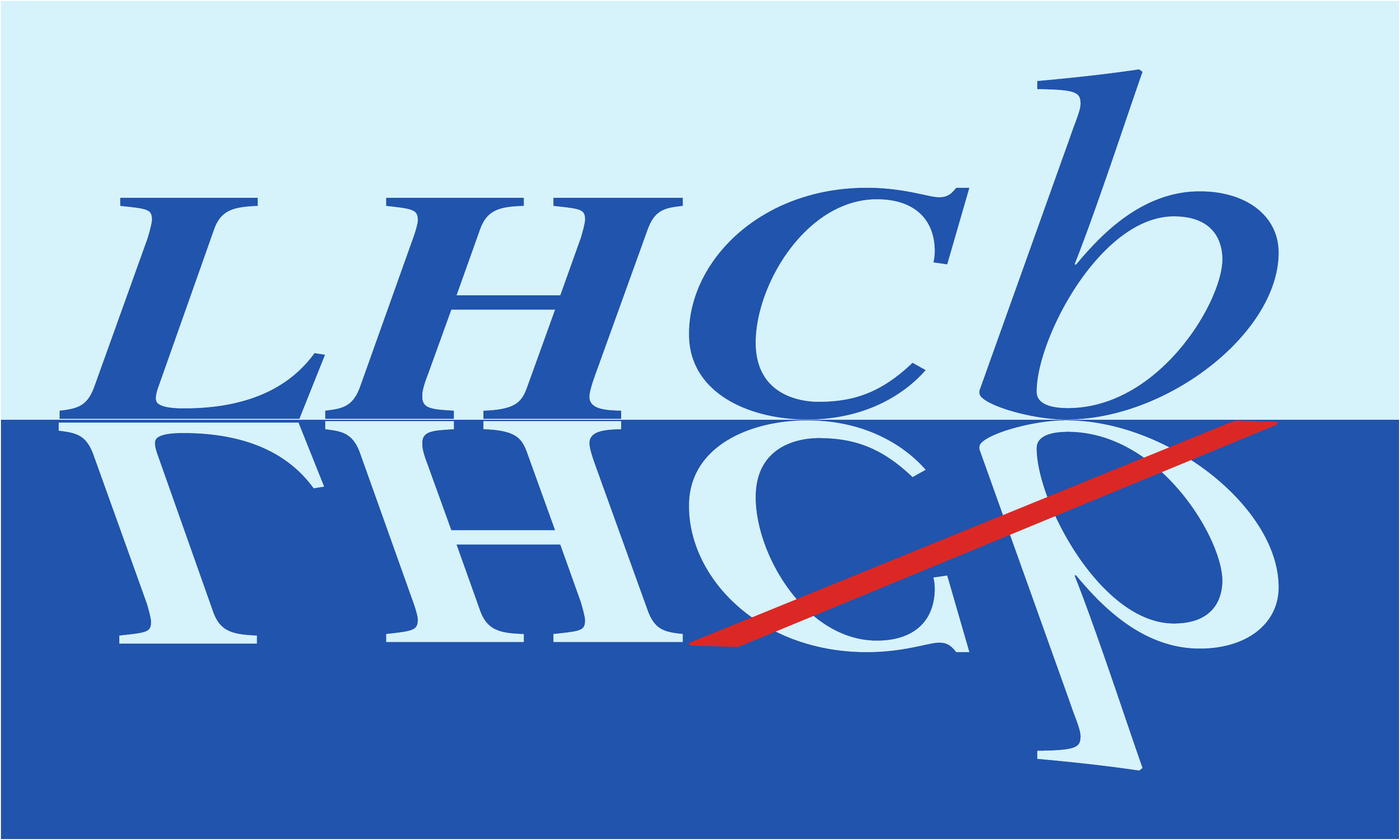}} & &}%
{\vspace*{-1.2cm}\mbox{\!\!\!\includegraphics[width=.12\textwidth]{LHCb/Figures/lhcb-logo.eps}} & &}%
\\
 & & CERN-EP-2026-234 \\  
 & & LHCb-PAPER-2026-028 \\  
 & & 2 September 2026 \\ 
 & & \\
\end{tabular*}

\vspace*{2.0cm}

{\normalfont\bfseries\boldmath\huge
\begin{center}
  \papertitle 
\end{center}
}

\vspace*{2.0cm}

\begin{center}
\paperauthors\footnote{Authors are listed at the end of this Letter.}
\end{center}

\vspace{\fill}

\begin{abstract}
  \noindent
  Tests of fundamental symmetries, such as $CPT$ and Lorentz invariance, provide sensitive probes of physics beyond the standard quantum-field-theory framework. Possible violations of these symmetries are investigated through interference effects in decays of $B_s^0$ mesons to $D_s^-\pi^{+}$ states. The analysis uses proton-proton collision data collected with the LHCb detector in 2015--2018 at a centre-of-mass energy of $13~\mathrm{TeV}$, corresponding to an integrated luminosity of $6~\mathrm{fb}^{-1}$. The results are interpreted within the Standard Model Extension framework, yielding sensitivities to the Lorentz-violating parameter $\Delta a_{\mu}$ at the ${\cal O}(10^{-14})\,\mathrm{GeV}$ level. Without assuming Lorentz-invariance violation, the $C\!PT$-violating parameter $z$ in the $B_s^0$ system is measured to be $\mathcal{R}e(z) = 0.018 \pm 0.016 \pm 0.003$ and $\mathcal{I}m(z) = 0.003 \pm 0.004 \pm 0.002$, where the first uncertainties are statistical and the second systematic. These results represent the most precise constraints on $C\!PT$ violation in the $B_s^0$ system. 
\end{abstract}

\vspace*{2.0cm}

\begin{center}
  Submitted to Phys. Rev. Lett. 
\end{center}

\vspace{\fill}

{\footnotesize 
\centerline{\copyright~\papercopyright. \href{\paperlicenceurl}{\paperlicence}.}}
\vspace*{2mm}

\end{titlepage}


\newpage
\setcounter{page}{2}
\mbox{~}
%
%
%
%


\renewcommand{\thefootnote}{\arabic{footnote}}
\setcounter{footnote}{0}


\cleardoublepage

\pagestyle{plain} 
\setcounter{page}{1}
\pagenumbering{arabic}


Lorentz invariance and the combined symmetry of charge conjugation, parity, and time reversal (\CPT) are fundamental to local relativistic quantum field theory~\cite{Greenberg:2002uu}. Several proposed approaches to quantum gravity suggest that these symmetries may be broken at the Planck scale ($10^{19}\gev$)~\cite{PhysRevD.39.683, Kostelecky:1991ak, Kostelecky:1994rn}. High-precision measurements at experimentally accessible energy scales can provide sensitivity to the tiny effects predicted by such approaches~\cite{Mavromatos_2004}.

Neutral meson systems provide a highly sensitive laboratory to probe such phenomena. The interference between the meson-antimeson mixing and decay processes can amplify even tiny departures from \CPT symmetry~\cite{Kostelecky:1997mh, Liberati:2013xla, particles7030042}. Within the framework of the Standard Model Extension (SME)~\cite{Colladay:1996iz, Colladay:1998fq, PhysRevD.69.105009}, potential Lorentz- and \CPT-violating effects are parameterized as couplings between conventional matter fields and fixed background tensors~\cite{Casimireffect}.

In the case of the \Bs--\Bsb system, \CPT violation would appear in the time evolution of the two flavor eigenstates. The dynamics are governed by the \schr equation~\cite{PDG2026}
\begin{align}
i \partial_t \Psi = \hat{H} \Psi ,
\end{align}
where the $2\times2$ effective Hamiltonian, $\hat{H} = \hat{M} - i\hat{\Gamma}/2$, describes \Bs mixing~\cite{PDG2026}. Diagonalization of $\hat{H}$ defines the heavy \BsH and light \BsL mass eigenstates with masses \mHL and decay widths \GHL, leading to the observables $\dms \equiv \msH - \msL$ and $\DGs \equiv \GsL - \GsH$~\cite{Aubert_2004, HFLAV:2024ctg}. A difference between the diagonal elements of $\hat{M}$ or $\hat{\Gamma}$ in $\hat{H}$ would indicate \CPT violation, and can be expressed through the complex parameter~\cite{Kosteleck_2001}
\begin{align}
z \equiv \frac{\delatms - i\dGs/2}{\dms + i\DGs/2} \ ,
\end{align}
where $\delatms \equiv M_{11}-M_{22}$ and $\dGs\equiv \Gamma_{11}-\Gamma_{22}$ denote the differences in the  diagonal mass and decay-width terms, respectively. 

Considering only terms up to first order in the parameter $z$, the decay-time distribution of mesons produced as \Bs or \Bsb flavor eigenstates in proton--proton (\proton\proton) collisions is given in Ref.~\cite{LHCb-PAPER-2016-005}. The initial flavor of the \Bs meson can be determined experimentally using flavor-tagging algorithms. For flavor-specific decays such as \BsToDsPim,\footnote{Charge-conjugate processes are implied unless otherwise stated.} the parameter describing \CP violation in the interference between mixing and decay satisfies $S_{\Dsmpip}=0$, while the parameter relating to direct \CP violation satisfies $C_{\Dsmpip} = 1$. The decay-time distribution can then be written as
\begin{align}
P_{\Bs(\Bsb)}(t) \propto e^{-\Gs t}\Big[&
\cosh(\DGs t/2)
+ \Rez [C_{\Dsmpip} + \qf] \sinh(\DGs t/2) \nonumber\\
& + \qf C_{\Dsmpip}\,\cos(\dms t)
+ \qf \Imz [C_{\Dsmpip} + \qf ] \sin(\dms t)
\Big],
\label{eq:mastercp}
\end{align}
where $\qf=+1$ ($-1$) for an initially produced \Bs (\Bsb) meson.

A nonzero \Rez introduces a $\sinh(\DGs t/2)$ term into the decay-time distribution, while a nonzero \Imz introduces a $\sin(\dms t)$ term. The presence of these terms constitutes the experimental signature of \CPT violation in this measurement.

As $\dmd \ll \dms$ and \DGd is negligible, \Bd mesons are predominantly sensitive to \Imz, while sensitivity to \Rez is strongly suppressed~\cite{Apostolakis:386076, Higuchi_2012,PhysRevD.67.052004,PhysRevLett.100.131802,LHCb-PAPER-2016-005}. The \Bs system therefore provides an important complementary probe, as its sizeable decay-width difference gives access to \Rez. Together, measurements in different neutral-meson systems also probe the dependence of the coefficients governing \CPT and Lorentz violation on the valence-quark content of the meson~\cite{LHCb-PAPER-2016-005,201489,LINK20037,PhysRevLett.100.131802,PhysRevD.82.101702}. Within the SME, \CPT- and Lorentz-violating effects in neutral-meson mixing can be parameterized by a real four-vector coefficient, denoted \Damu, which explicitly characterizes Lorentz-invariance violation. This coefficient couples to the meson four-velocity $\beta^{\mu} = \gamma(1,\vec{\beta})$, yielding~\cite{Kosteleck_2001, LHCb-PAPER-2016-005}
\begin{align}
  z = \frac{\beta^{\mu}\Damu}{\dms + i\DGs/2}\,.
  \label{eq:zdamu}
\end{align}
Since $\dms \gg \DGs$~\cite{PDG2026}, the contribution from \Imz to the determination of \Damu is strongly suppressed and can therefore be neglected without affecting the sensitivity of the measurement. The large mean boost of \Bs mesons in the \lhcb acceptance ($\langle \gamma\beta \rangle \approx 20$) enhances the sensitivity to these effects~\cite{vanTilburg:2014dka, LHCb-PAPER-2016-005}.

To account for the Earth’s rotation, the meson momentum is expressed in a sidereal reference frame defined with respect to distant stars~\cite{Kostelecky:1999bm,Bluhm:2003un,Kostelecky:2008ts}. Since  $\Damu$ is defined in this inertial frame, the orientation of the meson momentum relative to $\Damu$ changes with sidereal time, resulting in a characteristic sidereal modulation of the measured signal. In this frame, the $Z$-axis is aligned with the Earth’s rotation axis and the $X$-axis points toward the vernal equinox. For \Bs mesons produced approximately along the beam axis, the real part of $z$ can be written as~\cite{LHCb-PAPER-2016-005}
\begin{align}
\Rez
&= \frac{\dms}{\dms^2+\DGs^2/4}\,\beta^\mu\Damu \nonumber\\
&= \frac{\gamma\dms}{\dms^2+\DGs^2/4}
\Big[
\Daze + \cos\chi\,\Daz
+ \sin\chi\big(
\Day\sin\phi_\text{sid} + \Dax\cos\phi_\text{sid}
\big)
\Big],
\label{eq:siddamu}
\end{align}
where $\chi=112.4^\circ$ is the angle between the beam axis and the Earth’s rotation axis, and $\phi_\text{sid}$ denotes the sidereal phase~\cite{LHCb-PAPER-2016-005}.

The measurement is most sensitive to the parameter combination parallel to the beam direction,
\begin{equation}
\Dap \equiv \Daze + \cos\chi\,\Daz \approx \Dat,
\end{equation}
while the orthogonal component,
\begin{equation}
\Dav \equiv -\cos\chi\,\Daze + \Daz \approx \Das,
\end{equation}
is probed with substantially reduced sensitivity due to the small transverse component of the \Bs velocity. Both components are measured and their correlation is found to be negligible.

Previous experimental studies of \CPT and Lorentz-invariance violation in the \Bs system are scarce. The only dedicated measurement was performed by \lhcb using \BsToJPsiPhi decays with a data sample corresponding to 3\invfb~\cite{LHCb-PAPER-2016-005}. Constraints have also been obtained by the \dzero collaboration~\cite{PhysRevLett.115.161601}.

This Letter presents a search for \CPT- and Lorentz-invariance violation in \BsToDsPip decays using a decay-time-dependent analysis. The decay is flavor-specific (the corresponding wrong-sign process \BsToDsPim is highly suppressed) and direct \CP violation is expected to be negligible~\cite{Gershon_2022, LHCb-PAPER-2025-074}. The procedure used to determine the values of \Rez, \Imz, and \Damu follows Ref.~\cite{LHCb-PAPER-2016-005}. For the \CPT-violation measurement, the parameter $z$ is taken to be constant, whereas for the Lorentz-invariance-violation search it is allowed to vary with time and with the kinematic properties of the \Bs meson. The analysis uses \proton\proton collision data collected with the \lhcb detector during Run~2 of the \lhc (2015–2018) at a center-of-mass energy of 13\tev, and corresponding to an integrated luminosity of 6\invfb.

The \lhcb detector~\cite{LHCb-DP-2008-001,LHCb-DP-2014-002} is a single-arm forward spectrometer covering the pseudorapidity range $2<\eta<5$, designed for precision studies of heavy-flavor hadrons containing \bquark or \cquark quarks. Of particular relevance to this analysis are the tracking and particle-identification (PID) systems. The tracking system comprises a silicon-strip vertex detector (VELO) surrounding the \proton\proton interaction region, a large-area silicon-strip detector located upstream of a $4{\mathrm{\,T\,m}}$ dipole magnet, and a combination of silicon-strip and straw drift-tube stations downstream, providing excellent momentum and vertex resolution~\cite{LHCb-DP-2013-003,LHCb-DP-2017-001}. Particle identification is achieved using two Ring-Imaging Cherenkov (\rich) detectors~\cite{LHCb-DP-2012-003}, whose performance in simulated events is calibrated using control samples~\cite{LHCb-PUB-2016-021,LHCb-DP-2019-001}. Candidate events are selected by a two-level trigger system~\cite{LHCb-DP-2012-004,LHCb-DP-2019-001}. At the hardware trigger level, events are required to have a muon with high transverse momenta~(\pt) or a hadron, photon or electron with high transverse energy in the calorimeters. The software trigger requires a two-, three- or four-track secondary vertex with a significant displacement from any primary \proton\proton interaction vertex~(PV). At least one charged particle must have high \pt and be inconsistent with originating from a PV. A multivariate algorithm~\cite{Roe,BBDT,LHCb-PROC-2015-018} is used to identify secondary vertices consistent with the decay of a \bquark hadron. The software trigger decision is required to be caused by tracks from the decay of the signal \Bs candidate. Simulated samples of \BsToDsPip decays are used to model the detector acceptance, reconstruction efficiency, and signal mass shapes. These samples are generated using \pythia~\cite{Sjostrand:2007gs} to simulate \proton\proton collisions, \evtgen~\cite{Lange:2001uf} for hadron decays, \photos~\cite{Golonka:2005pn} to account for final-state radiation, and the \geant toolkit~\cite{Allison:2006ve,*Agostinelli:2002hh} to describe the detector response, as configured in Ref.~\cite{LHCb-PROC-2011-006}. The simulated signal samples are further calibrated using various control channels to account for residual differences between data and simulation~\cite{LHCb-DP-2018-001}.

The \Bs mesons are produced with large momenta and typically travel about 1\cm before decaying. Their decay products form a distinct secondary vertex, which is significantly displaced from the PV, and have higher \pt than most other particles in the event~\cite{LHCb-PAPER-2021-005, LHCb-PAPER-2025-001}. The candidate selection, detector modelling, and signal extraction for \BsToDsPip decays follow Refs.~\cite{LHCb-PAPER-2021-005, LHCb-PAPER-2024-020}, exploiting these distinctive topological and kinematic features. In the final reconstruction, \Bs candidates are formed by combining \Dsm mesons decaying into \KmKpPim and \PimPipPim final states, with a companion pion. Additional requirements on track quality, vertex displacement, and momentum together with a dedicated multivariate classifier trained on simulated samples, further enhance signal discrimination. The full selection achieves a signal purity of approximately 84\%, with the selection optimized using simulation to maximize the expected sensitivity to the \CPT- and Lorentz-violating parameters by balancing signal efficiency and statistical precision. To avoid experimenter's bias, the values of the \CPT- and Lorentz-violating parameters were not examined until the analysis procedure had been finalized.

The selected \BsToDsPip candidates are analyzed through a simultaneous unbinned two-dimensional likelihood fit to the \Dsmpip and \KmKpPim or \PimPipPim mass distributions, performed separately for each data-taking period (2015–2016, 2017, 2018). This fit determines the signal yields and uses the \sPlot technique~\cite{Pivk:2004ty} to assign per-candidate background-subtraction weights for the subsequent decay-time analysis. After the full selection, four categories of background remain: random track combinations forming combinatorial background; partially reconstructed \BsToDsstPi decays where the photon from \DssmToDsgamma is not reconstructed; cross-feed from other \bquark hadrons producing the same final state, \BdToDsPi; and misidentified decays with similar final-state topologies, including \BdToDmpPipm, \LbToLcPib, and \BsToDsK. The latter are further reduced through dedicated kinematic vetoes and refined PID requirements.

\begin{figure}[t]
    \centering
    \includegraphics[width=\textwidth]{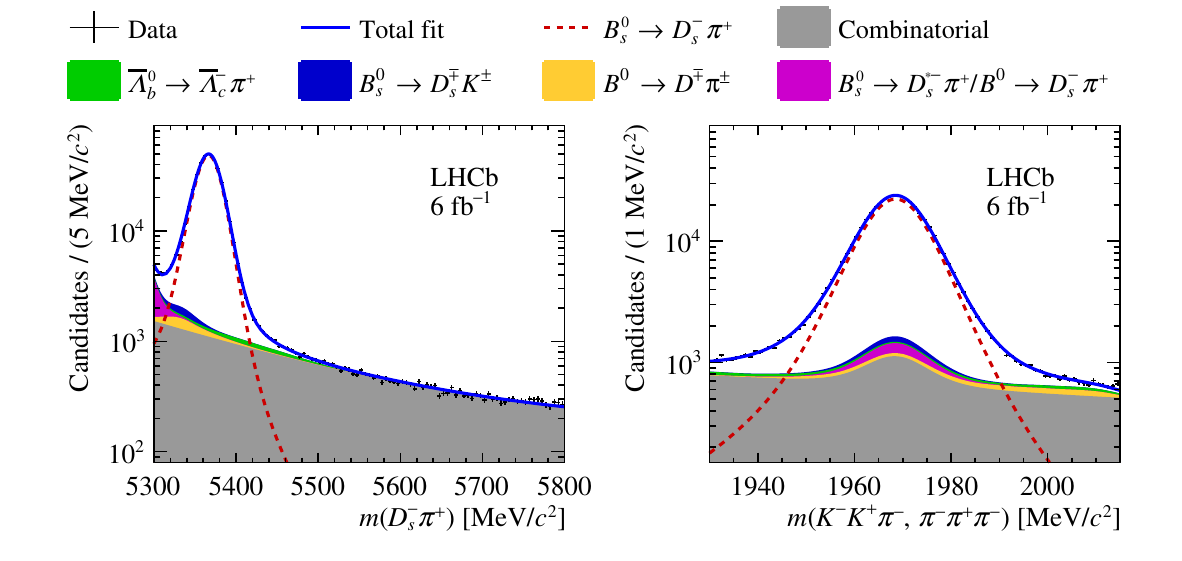}
    \caption{Mass distributions with fit results of the selected \BsToDsPip candidates, (left)~$m(\Dsmpip)$ and (right)~$m(\KmKpPim,\PimPipPim)$, combining all reconstructed \Dsm final states and the three data-taking periods.}
    \label{fig:massfit}
\end{figure}

The shapes of the signal and background mass components are obtained from simulated samples. The $m(\Dsmpip)$ and $m(\KmKpPim)$ or $m(\PimPipPim)$ signal peaks are modeled as the sum of a Hypatia~\cite{Santos:2013gra} and a Johnson $S_U$~\cite{JohnsonSU} function. The combinatorial background in the $m(\Dsmpip)$ distribution is described by the sum of two exponential functions, one of which has a smaller slope fixed using the high-mass sideband region, \mbox{$5600 < m(\Dsmpip) < 6800\mevcc$}. For the $m(\KmKpPim)$ or $m(\PimPipPim)$ distributions, the background is modeled by an exponential component describing random track combinations together with a peaking component arising from \Dsm mesons. Contributions from \BdToDmpPipm, \LbToLcPib, and \BsToDsK decays are modeled using shapes taken from simulation, with no free parameters in the fit. Their yields are constrained relative to the signal using known branching-fraction ratios~\cite{PDG2026}, efficiency ratios from simulation, and corrections for residual data-simulation differences. Backgrounds from \BdToDsPi and \BsToDsstPi decays are taken from simulation with their combined yield free to vary in the fit. The resulting mass spectra combining all reconstructed \Dsm final states and the three data-taking periods are shown in Fig.~\ref{fig:massfit}, from which a signal yield of \mbox{$378\,500 \pm 700$} is measured.

The probability density function describing the decay-time distribution is modified to account for detector-related effects, including a decay-time-dependent reconstruction efficiency, finite decay-time resolution and possible biases, imperfect determination of the initial \Bs or \Bsb flavor, production asymmetries
in \proton\proton collisions, and charge-dependent reconstruction asymmetries arising from interactions of the final-state particles with the detector material~\cite{LHCb-PAPER-2016-062, LHCb-PAPER-2023-001}.

Since the \CPT- and Lorentz-invariance-violating parameters depend on the decay-time-dependent reconstruction efficiency, its accurate determination is essential~\cite{PhysRevLett.115.161601, vanTilburg:2014dka}. This efficiency is determined using a flavor-specific control channel. For \BdToDPi decays, it is extracted from a fit to the decay-time distribution, with the \CP-violation parameters and time-evolution constants, \Gd, \DGd and \dmd, fixed to their known values~\cite{PDG2026} and is parameterized using cubic B-spline segments~\cite{LHCb-PAPER-2019-013}, implemented analytically within the decay-time fit~\cite{Karbach:ComplxErrFunc}. The resulting efficiency is then used in the measurement of \CPT- and Lorentz-invariance-violating parameters in \BsToDsPip decays, and corrected by the ratio of decay-time acceptances between the two modes, which is determined from simulation~\cite{LHCb-PAPER-2024-020}. The number of spline segments is optimized to describe the efficiency at low decay times. In the decay-time probability density function, the efficiency function enters as a multiplicative factor in the signal model.

The decay-time resolution is evaluated using a sample of promptly produced \Dsm mesons in \proton\proton collisions, not originating from a \Bs meson decay. Candidates are selected with the same trigger selection as additional offline selection requirements to obtain a high-purity \Dsm sample without requiring displacement from the PV. The multivariate classifier employed in the signal selection is not applied. The decay time is reconstructed from the distance between the \Dsm production vertex and an artificial \Bs decay vertex, formed by combining the \Dsm candidates with a randomly chosen \pip from the same event, and is expected to be zero for prompt decays up to resolution and detector biases. A linear relation between the true resolution and per-candidate uncertainty from the vertex fit is assumed in calibration of the decay-time uncertainty in \BsToDsPip decays.

A small bias in the reconstructed \Bs decay time, originating from residual detector misalignments, was previously identified in the measurement of \dms using \BsToDsPip decays~\cite{LHCb-PAPER-2021-005}. This bias is quantified using a control sample of prompt \Dsm mesons, with corrections determined by comparison with simulated samples that are intentionally misaligned. The present analysis applies the same correction, adjusted to account for differences in the event selection. The correction is determined separately for each data-taking period and corresponds to approximately 0.005\ps~\cite{LHCb-PAPER-2021-005}.

Determining the initial flavor of the \Bs meson is crucial for measuring its time-dependent decay rate. Flavor-tagging algorithms exploit properties of \bquark-hadron production in \proton\proton collisions. Decays of the opposite-side~(OS) \bquark hadron provide information on the signal \Bs flavor as \bquark and \bquarkbar quarks are produced in pairs~\cite{LHCb-PAPER-2011-027}. Muon and electron taggers target semileptonic \bquark decays, while kaon and charm taggers use the chains $\bquark \rightarrow \cquark \rightarrow \squark$ and $\bquark \rightarrow \cquark$, respectively. The vertex-charge tagger reconstructs an effective charge from a displaced vertex of the accompanying \bquark decay~\cite{LHCb-PAPER-2015-027}. Complementing these, same-side (SS) kaon taggers identify the charge of kaons produced in the \squark\squarkbar fragmentation pair to infer the initial \Bs-meson flavor~\cite{LHCb-PAPER-2015-056}.

Each tagger provides a flavor decision and an estimated mistag probability \Peta, which may differ from the true probability \Pomega in data. For OS taggers, \Peta is calibrated using a control sample of \BuToJpisKp decays~\cite{LHCb-PAPER-2021-005} via the linear relation $\Pomega = p_0 + p_1 \Peta$, with the $p_0$ and $p_1$ parameters determined separately for each data-taking period. The combined OS decision and mistag estimate is obtained following the procedure in Ref.~\cite{LHCb-PAPER-2011-027}. 

To determine the values of \Rez, \Imz, and \Damu, two unbinned maximum-likelihood fits are performed over all reconstructed \Dsm final states and the three data-taking periods. For each period, the decay-time resolution and mistag-probability calibrations are constrained to the values corresponding to that period. The fitted covariance matrix is corrected using the asymptotically valid procedure described in Ref.~\cite{Langenbruch_2022}, ensuring proper uncertainty coverage in the presence of event weights from the \sPlot technique. The decay-time fits are performed using Eq.~\ref{eq:mastercp}.

In both fits, the calibration parameters of the SS kaon taggers and the flavor-tagging mistag-probability asymmetries for both OS and SS taggers are allowed to vary. The parameters \Gs, \DGs, and \dms are fixed to their known values~\cite{PDG2026}. Other fixed inputs include: $C_{\Dsmpip}$, as deviations are expected to be negligible at the current experimental precision~\cite{LHCb-PAPER-2021-005,LHCb-PAPER-2024-020}; the estimated wrong-tag fractions and efficiencies of the OS and SS taggers; the calibration parameters of the OS taggers; the decay-time-dependent reconstruction efficiency; the decay-time bias correction; and the decay-time resolution calibration. The \Bs--\Bsb production and detection asymmetries are fixed to zero, as they are negligible at the current experimental precision~\cite{LHCb-PAPER-2016-062, LHCb-PAPER-2025-074}.

\begin{figure}[b]
  \begin{center}
    \includegraphics[scale=0.38]{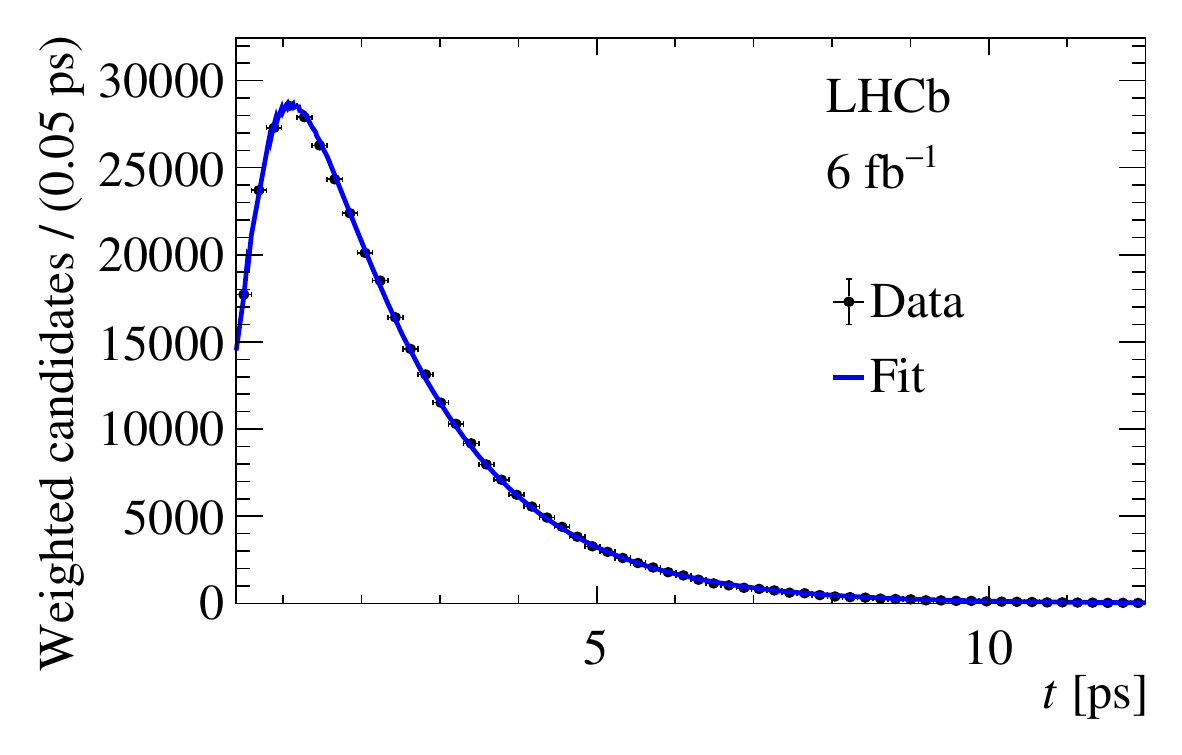}
    \includegraphics[scale=0.38]{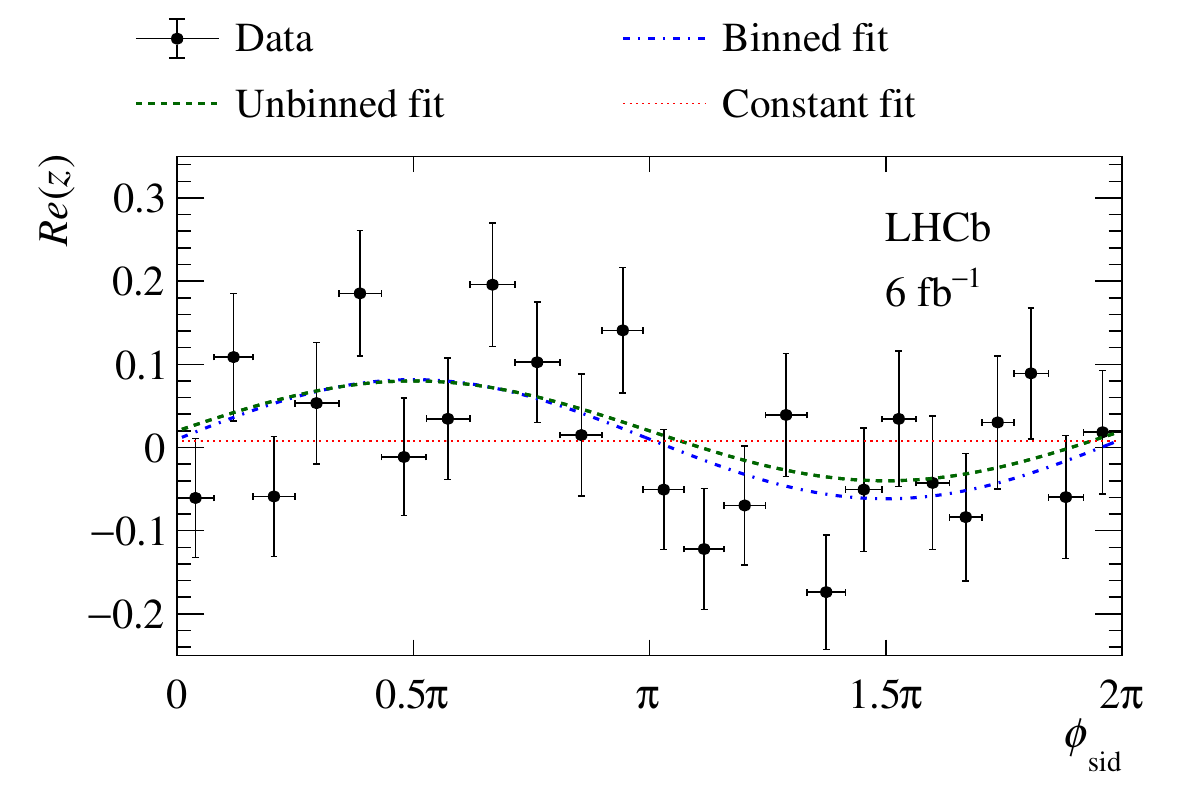}
  \end{center}
  \caption{(Left) Decay-time distribution of \BsToDsPip signal candidates, with the fit result overlaid. (Right)~Sidereal-phase dependence of \Rez, where the solid blue curve shows the binned fit using Eq.~\ref{eq:siddamu}, while the dashed green curve represents the prediction from the unbinned fit to the full dataset, evaluated with the same average \Bs momentum. The dashed red curve corresponds to a fit to the 24 sidereal-phase intervals under the hypothesis of no sidereal-phase dependence.}
\label{fig:sfit}
\end{figure}

In the first fit, the \CPT-violating parameters \Rez and \Imz are determined. The decay-time spectrum for the full dataset is shown in Fig.~\ref{fig:sfit}. In the second fit, \Rez in Eq.~\ref{eq:mastercp} is parameterized in terms of the Lorentz-invariance-violating coefficients \Damu using Eq.~\ref{eq:siddamu}, and the parameters \Damu are extracted with \Imz fixed to zero, as discussed earlier.

To validate the \Damu coefficients obtained from the unbinned likelihood fit, the full dataset is divided into 24 intervals of sidereal phase. A simultaneous fit is performed to extract \Rez in each bin without imposing any sidereal dependence. The resulting \Rez values are then described as a function of sidereal phase using the compact expression in Eq.~\ref{eq:siddamu}, assuming the average \Bs momentum. This provides an independent determination of the sidereal modulation induced by the \Damu coefficients. The result of the binned fit and the expectation obtained by inserting the \Damu value from the second unbinned fit into Eq.~\ref{eq:siddamu} using the same average \Bs momentum are shown in Fig.~\ref{fig:sfit}. The good agreement between the two confirms the consistency of the extracted \Damu coefficients.

Systematic uncertainties arise from assumptions in the signal and background models and from fixed fit parameters. Following the strategy of Ref.~\cite{LHCb-PAPER-2024-020}, these are grouped into independent sources. All uncertainties, except those related to fixed shapes and yields in the two-dimensional mass fit, are evaluated using ensembles of 1000 pseudoexperiments generated from baseline-fit results, varying relevant parameters within their uncertainties. The resulting shifts in fitted parameters quantify the systematic uncertainty, including both the peak position shift and its spread. Correlations between parameters, such as \Rez and \Imz or components of \Damu, are captured in a covariance matrix scaled to the total systematic uncertainty. Reconstruction-related contributions are dominated by the momentum-scale uncertainty, which affects the decay-time determination, and is evaluated by varying the simulated momenta. Effects from the VELO length scale and detector alignment are found to be negligible. The value of $\dms$ is varied within its known uncertainty~\cite{PDG2026}, while production and detection asymmetries are allowed to vary freely. Coherent variations of \Gs and \DGs follow their known correlation~\cite{Lyons:1988rp}. Alternative decay-time resolution models, spline variations in decay-time acceptance, and potential biases from neglected correlations between mass and decay time, evaluated with bootstrap sampling, are also considered. A summary of all systematic uncertainties is given in Table~\ref{tab:TotalSystErrSfit}.

In summary, the parameters governing potential \CPT and Lorentz-invariance violation in \BsToDsPip decays are determined using proton-proton collision data recorded by the \lhcb detector at a center-of-mass energy of 13\tev, corresponding to integrated luminosity of 6\invfb. The measured values are 
\begin{align*}
\Rez         &= \RezResult\ ,  \\ 
\Imz         &= \ImzResult\ ,  \\
\Dap         &= \DatResult\ ,  \\
\Dav         &= \DasResult\ ,  \\
\Dax         &= \DaxResult\ ,  \\
\Day         &= \DayResult\ ,  
\label{measure:result}
\end{align*}
where the first uncertainties are statistical and the second systematic.

The results for \Rez and \Imz are consistent with \CPT symmetry, and the measurement of \Damu show no evidence for Lorentz-invariance violation. This provides the most stringent constraints on both \CPT and Lorentz-invariance violation in the \Bs system. A comparison of the oscillatory and constant hypotheses using the 24 sidereal-phase intervals yields a significance of 1.8 standard deviations. This is compatible with a statistical fluctuation and provides no evidence for the sidereal modulation expected from Lorentz-invariance violation.
\section*{Acknowledgements}
%
%
\noindent We express our gratitude to our colleagues in the CERN
accelerator departments for the excellent performance of the LHC. We
thank the technical and administrative staff at the LHCb
institutes.
We acknowledge support from CERN and from the national agencies:
ARC (Australia);
CAPES, CNPq, FAPERJ and FINEP (Brazil); 
MOST and NSFC (China); 
CNRS/IN2P3 and CEA (France);  
BMFTR, DFG and MPG (Germany);
NKFIH (Hungary);              
INFN (Italy); 
NWO (Netherlands); 
MNiSW and NCN (Poland); 
MEC/IFA (Romania); 
MICIU and AEI (Spain);
SNSF and SER (Switzerland); 
NASU (Ukraine); 
STFC (United Kingdom); 
DOE NP and NSF (USA).
We acknowledge the computing resources that are provided by ARDC (Australia), 
CBPF (Brazil),
CERN, 
IHEP and LZU (China),
IN2P3 (France), 
KIT and DESY (Germany), 
INFN (Italy), 
SURF (Netherlands),
Polish WLCG (Poland),
IFIN-HH (Romania), 
PIC (Spain), CSCS (Switzerland), 
GridPP (United Kingdom),
and NSF (USA).  
We are indebted to the communities behind the multiple open-source
software packages on which we depend.
Individual groups or members have received support from
RTP (Australia), 
FWO Odysseus grant G0ASD25N (Belgium), 
Key Research Program of Frontier Sciences of CAS, CAS PIFI, CAS CCEPP (China); 
Minciencias (Colombia);
EPLANET, Marie Sk\l{}odowska-Curie Actions, ERC and NextGenerationEU (European Union);
A*MIDEX, ANR, IPhU and Labex P2IO, and R\'{e}gion Auvergne-Rh\^{o}ne-Alpes (France);
Alexander-von-Humboldt Foundation (Germany);
ICSC (Italy); 
Severo Ochoa and Mar\'ia de Maeztu Units of Excellence, GVA, XuntaGal, GENCAT, InTalent-Inditex and Prog.~Atracci\'on Talento CM (Spain);
the Leverhulme Trust, the Royal Society and UKRI (United Kingdom).

\clearpage
\section*{End matter}
\label{sec:appendix}
\appendix
\section{Summary of systematic uncertainties}
\label{sec:app:syst}
The systematic uncertainties on the measured parameters are summarized in Table~\ref{tab:TotalSystErrSfit}.  
\begin{table}[h]
\centering
\caption{\label{tab:TotalSystErrSfit} Systematic uncertainties contributing to the measured \CPT- and Lorentz-invariance-violating parameters. The symbol ``$-$" indicates a negligible contribution. The total systematic uncertainty represents the quadrature sum of all individual contributions.}
\resizebox{\textwidth}{!}{
\renewcommand{\arraystretch}{1.1}
\begin{tabular}{l|cc|rccc}
\hline \hline
Description & \Rez  & \Imz & \multicolumn{1}{c}{\Dap} & \Dav $\times 10^{2}$ & \Dax & \Day \\ 
& \multicolumn{2}{c|}{$\times 10^{-2}$} & \multicolumn{4}{c}{$\times 10^{-14}$ GeV}  \\ \hline
Two-dimensional mass model  & 0.10 & 0.04 & 0.04 & 0.03 & 0.16 & 0.17 \\
Momentum scale                & 0.08     & 0.01 & $-$ & 0.03 & 0.01 & 0.02 \\
Fixed \dms & 0.11 & 0.04 & 0.12 & 0.03 & 0.26  & 0.23     \\
Decay-time acceptance  & 0.13 & 0.07 & 0.21 & 0.02 & 0.29 & 0.27     \\
Fixed \Gs and \DGs  & 0.18 & 0.12 & 0.17 & 0.06 & 0.25 & 0.24 \\
Production and detection asymmetry & 0.07 & 0.02  & 0.07 & 0.02 & 0.10 & 0.11 \\
Decay-time resolution model   & 0.09 & 0.02 & 0.08 & 0.01 & 0.15 & 0.11     \\
Correlation among observables & 0.05 & 0.03 & 0.01 & 0.01 & 0.13 & 0.14     \\ \hline
Total                          & 0.31 & 0.16 & 0.32 & 0.07 & 0.53 & 0.51   \\ \hline
\end{tabular}
}
\end{table}
\newpage

\addcontentsline{toc}{section}{References}
\bibliographystyle{LHCb/LHCb}
\bibliography{main,LHCb/standard,LHCb/LHCb-PAPER,LHCb/LHCb-CONF,LHCb/LHCb-DP,LHCb/LHCb-TDR,LHCb/LHCb-PUB}

\newpage
\centerline
{\large\bf LHCb collaboration}
\begin
{flushleft}
\small
R.~Aaij$^{39}$\lhcborcid{0000-0003-0533-1952},
M.~Abdelfatah$^{71}$,
A.S.W.~Abdelmotteleb$^{59}$\lhcborcid{0000-0001-7905-0542},
C.~Abellan~Beteta$^{53}$\lhcborcid{0009-0009-0869-6798},
F.~Abudin\'en$^{61}$\lhcborcid{0000-0002-6737-3528},
T.~Ackernley$^{63}$\lhcborcid{0000-0002-5951-3498},
A.A.~Adefisoye$^{71}$\lhcborcid{0000-0003-2448-1550},
B.~Adeva$^{49}$\lhcborcid{0000-0001-9756-3712},
M.~Adinolfi$^{57}$\lhcborcid{0000-0002-1326-1264},
P.~Adlarson$^{87,44}$\lhcborcid{0000-0001-6280-3851}
% ...

{\footnotesize \it

% ...
$^{39}$Nikhef National Institute for Subatomic Physics, Amsterdam, Netherlands\\
% ...
$^{44}$National Center for Nuclear Research (NCBJ), Warsaw, Poland\\
% ...
$^{49}$Instituto Galego de F{\'\i}sica de Altas Enerx{\'\i}as (IGFAE), Universidade de Santiago de Compostela, Santiago de Compostela, Spain\\
% ...
$^{53}$Physik-Institut, Universit{\"a}t Z{\"u}rich, Z{\"u}rich, Switzerland\\
% ...
$^{57}$H.H. Wills Physics Laboratory, University of Bristol, Bristol, United Kingdom\\
% ...
$^{59}$Department of Physics, University of Warwick, Coventry, United Kingdom\\
% ...
$^{61}$School of Physics and Astronomy, University of Edinburgh, Edinburgh, United Kingdom\\
$^{62}$School of Physics and Astronomy, University of Glasgow, Glasgow, United Kingdom\\
$^{63}$Oliver Lodge Laboratory, University of Liverpool, Liverpool, United Kingdom\\
% ...
$^{71}$Syracuse University, Syracuse, NY, United States\\
% ...
$^{87}$Department of Physics and Astronomy, Uppsala University, Uppsala, Sweden, associated to $^{62}$\\
% ...
}
\end{flushleft}

\end{document}